-RESEARCH ARTICLE-

# Multimedia Respiratory Database (RespiratoryDatabase@TR): Auscultation Sounds and Chest X-rays

Gokhan Altan[1*], Yakup Kutlu[2], Yusuf Garbi[2], Adnan Özhan Pekmezci[3], Serkan Nural[3]

[1]Department of Informatics, Mustafa Kemal University, Hatay, Turkey
[2]Department of Computer Engineering Iskenderun Technical University, Hatay, Turkey
[3]Antakya State Hospital, Hatay, Turkey

**Abstract**
Auscultation is a method for diagnosis of especially internal medicine diseases such as cardiac, pulmonary and cardio-pulmonary by listening the internal sounds from the body parts. It is the simplest and the most common physical examination in the assessment processes of the clinical skills. In this study, the lung and heart sounds are recorded synchronously from left and right sides of posterior and anterior chest wall and back using two digital stethoscopes in Antakya State Hospital. The chest X-rays and the pulmonary function test variables and spirometric curves, the St. George respiratory questionnaire (SGRQ-C) are collected as multimedia and clinical functional analysis variables of the patients. The 4 channels of heart sounds are focused on aortic, pulmonary, tricuspid and mitral areas. The 12 channels of lung sounds are focused on upper lung, middle lung, lower lung and costophrenic angle areas of posterior and anterior sides of the chest. The recordings are validated and labelled by two pulmonologists evaluating the collected chest x-ray, PFT and auscultation sounds of the subjects. The database consists of 30 healthy subjects and 45 subjects with pulmonary diseases such as asthma, chronic obstructive pulmonary disease, bronchitis. The novelties of the database are the combination ability between auscultation sound results, chest X-ray and PFT; synchronously assessment capability of the lungs sounds; image processing based computerized analysis of the respiratory using chest X-ray and providing opportunity for improving analysis of both lung sounds and heart sounds on pulmonary and cardiac diseases.

[*]*Corresponding author: Gökhan Altan, e-mail: gokhan_altan@hotmail.com*

**Introduction**

A computer-assisted medical interface (CMI) is frequently used method in today's technology for performing clinical procedures, applying various clinical tests to patient, interpreting and analysing the clinical tests, and transferring detailed and characteristic information in line with clinicians' expertise. In addition to its competencies, the development of the CMI has focused on generation of a large patient cloud by modernizing in order to find patients and to hide laboratory tests and to easily access the patients' specific clinical data all over the place. The CMI has great precautions in terms of the patient's ability to access clinical history, used drugs, X-rays, Electrocardiography (ECG) and the auscultation sounds such as lung and heart sounds to direct physicians to control different diseases simultaneously. The CMI provides the integration of methods that enable a physician and a patient to interact with a machine, such as medical device, software, or complex device. Nowadays, the usage of the CMI is not limited to physicians; also it allows communicating with other clinical personnel, diagnosis process with time-efficient, treatment and monitoring of the diseases using various sensors on mobile devices such as smart phones, tablets and wearable devices on the local and wide area networks.

     While some clinical data can be acquired via elaborated monitoring tools using wearable devices with sensible sensors, many biomedical signal acquisition and detection areas still require expert clinical physicians. Significant researches in medical areas should be supported to reduce the commitment to clinical experts during diagnosis and treatment monitoring processes. The assessment of auscultation sounds, such as heart sounds and lung sounds, requires characteristic and complex information accumulation that cannot be patient-self applying. Especially, the cases such as variety of the auscultation points according to the physical characteristics of the patients and auscultating the posterior lung sounds prevent the self-recording. The auscultation requires an expert clinician or patient-specific designed wearable devices.

     The medicine budget is the vast majority part in the treatment process of the diseases. In social innovation studies in the field of health, the treatment methods and studies on the pharmaceutical industry were conducted between 1980 and 2000; particularly more on vision and diagnostic systems while disease prevention, wearable health services were researched between 2000 and 2015. Nowadays, a fast and successful path in technology is a fact that many innovations are supposed to shape innovations for health. In the coming years, solutions for advanced wearable technologies, telemedicine platforms, mobile health platforms and patient management will have a large part of the diagnosis, monitoring and treatment stages of the diseases. The creation of such innovations on diseases has the greatest share in the transfer of digital data that will be collected about several diseases.

     One of the most important factors in the health policy is creating a healthy society, preventing the diseases, and keeping the disorders under control. According to World Health

Organization (WHO) reports, there are three of respiratory diseases in the deadliest list (WHO, 2015). Respiratory diseases occur due to environmental factors such as air pollution, hereditary causes, tobacco use or exposure to smoke, age, sex, race, infections, seasonal factors, environmental conditions and occupational factors. The most common respiratory disorders are asthma, bronchitis, chronic obstructive pulmonary disease (COPD), lower respiratory tract infections, alveolitis, pulmonary fibrosis, pneumonia, asbestosis, chronic bronchitis, bronchiectasis, laryngitis, anatomical hypothesis, laryngeal disturbances, airway inflammation, tumour and tracheal narrowing. While some of these diseases only accelerate the end-stage attacks as a result of seasonal changes, some are genetic disorders that are not treated, but can be kept under control (Troosters et al., 2005). Considering update on the WHO report in 2015, lower respiratory tract infection such as bronchitis, pneumonia takes places in the first rows with 3.20 million deaths and the a frequency of 291 million cases and the COPD takes the second position with 3.19 and a frequency of 174.5 million cases (GBD, 2016) in the respiratory diseases. Reaching such high levels of the respiratory diseases demonstrates the needs for research, diagnosis, and early diagnosis and treatment systems on this expert. Although it has ability to treat lower respiratory tract infections, no cure has been developed yet for the COPD, but with treatment it's preventable and manageable. Thus, early detection of the COPD and early stage control of the disease have a great impact on reducing death incidents due to respiratory diseases. The COPD is a disturbance that is seen in the people over 35 years old, especially due to the use of tobacco products, genetic factors or environmental factors. As a consequence of the COPD, the lungs lose their elasticity, thickening and inflammation in the airway layer, and it results increasing form of breathing shortness, periodic coughing crises and wheezing (Decramer et al., 2012). It usually possesses advanced forms of both the asthmatic and bronchitis symptoms together. Hence, the COPD is not a disease described by a single symptom. It has 5 levels (COPD0, COPD1, COPD2, COPD3 and COPD4) in the diagnosis processes. Assessment of clinical symptoms such as various respiratory tests, effort tests, pulmonary function test (PFT), X-rays, clinical examinations, chest examination, and other symptoms such as vital signs, clinical tests are crucial in the facilitating diagnosis of respiratory diseases as it is related to the patient's self-management skills (Troosters et al., 2005). After chest and back auscultation by chest specialists, there may be a need for questionnaires for assessment of impacts on patient's condition such as the environment (SGRQ-C), PFT, chest x-ray, tobacco history and different assays with some special conditions in the definitive diagnosis of the COPD. The respiratory diseases need to be improved in signal analysis researches to ensure the exact recognition on quickly, and regardless of the tests assessments.

The remaining of this paper is organized as follows; in materials and methods section, data acquisition stage, detailed description on clinical data, tests and the auscultation are given. Besides, a CMI and a multimedia database on the respiratory diseases are evaluated. The advantages of the database are presented in the result section.

**Materials and Methods**

The biggest deficiency in the field of respiratory disorders is in the course of data acquisition, benchmarking and evaluation stages. The diagnosis and detection analysis on the biomedical signals have many databases such as open-access ECG databases that describe heart conditions, cardiac abnormalities, specific EEG databases that are evaluated on analysis of neurological disorder, image databases that demonstrate the dermatological and pattern abnormalities and the respiratory sounds in commercial use that are based on wheezing and respiratory diseases. This is an obstacle for researchers in developing diagnosis and assessment processes of respiratory diseases and pulmonary based cardiac diseases. Because of the respiratory databases contain a restricted number of diseases; there are a limited number of studies on respiratory diseases, and respiratory sounds cannot be analysed for newly deadliest pulmonary diseases using newly developed and efficient transformation algorithms. In the light of mentioned deficiencies and limitations on the respiratory databases, it is a necessity that creating new respiratory databases that are based on the most deadliest diseases the COPD and lower respiratory tract diseases, and advanced levels of the COPD to ensure signal processing assessments and abnormality detections. It will be an excellent step for health care, upgrading the medical software with the ability to analyse the sounds by adding an effective and functional graphical user interface, the ability to label specific abnormalities parts such as crackle and wheezing on the respiratory sounds, the detailing of auscultation sounds and imaging reports, the visualization of auscultation sounds and the regional amplification on respiratory sounds for new database with ease of use. In this way, the respiratory history of patients supported by all different clinical tests can be maintained in one database, the course of further diagnostic and monitoring of the disease will be benefit greatly for clinicians and patients.

People have metabolisms that react differently according to the diseases and treatment processes. This specifications cause different responses to different treatment methods, different durations of procedures, conditions and treatments for patients with respiratory disorders. Also respiratory disease is a medical unit with symptoms that can vary for each patient. The acquired data in the basis of this information will positively affect the treatment process of the patients and it will be possible to control some abnormalities without experiencing attacks in early stages. Especially respiratory diseases in particular cause, the patient experiences severe attacks at certain stages and during seasonal transitions. The lung auscultation sounds, abnormalities wheezing and crackles during respiratory attacks, clinical variability, PFT variability, chest X-rays, and assessments of questionnaires will provide an advantage for clinicians and pulmonologists in the monitoring, early managing and regimen processes of the patient specific treatment. These advantages will result in a significant difference in the quality of durational differences and evaluation on the treatment and diagnosis of the chronic diseases. Such a CMI system, which can be accessed online, saves time in evaluations made by different pulmonologists, as well as it reduces national economic savings on medicine, use of diagnostic tools such as chest X-rays, clinical tests, PFT, and more. The data acquisition system is seen in Figure 1.

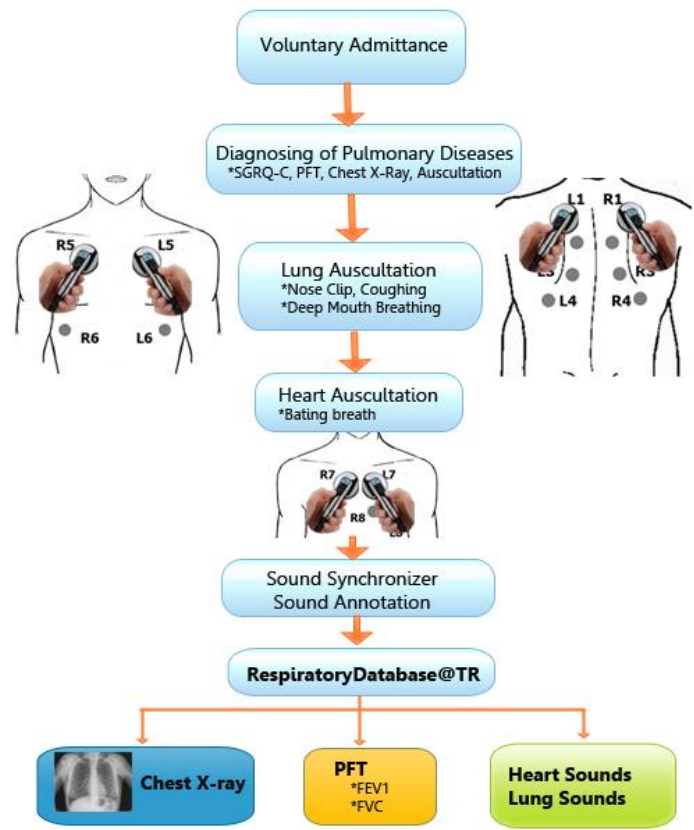

Figure 1. Data acquisition system of the RespiratoryDatabase@TR

*Chest X-Ray*

X-ray (Radiography) is a diagnostic tool that allows the human body to visualize patterns related to the structure of internal organs, tissues and bones by applying low amounts of radiation. Chest X-ray is taken lung-centred position and point's abnormalities on the lungs, obstructions in the airways, structure of the thoracic cusps, lungs and heart structure, blood vessels obstruction. Chest X-ray can also identify obstructions in the lungs, fluid concentration, inflammation (Friis et al., 1990). The chest X-ray is an easy, fast and effective way for doctors to control the condition of internal organs (Hederos et al., 2004). The designed CMI has a module which the patients' chest X-rays can be imported. The advanced toolbars that can adjust and contrast the X-rays have been added, especially in the labelling of patients in the display of the lungs and heart pattern. The designed CMI provides clearly displaying the obstructions and constrictions in the airways, the status of heart vessels such as embolism and fluid accumulation in the chest area using chest X-rays for diagnosis and managing of the pulmonary and cardiac diseases.

*Pulmonary Function Test*

A PFT is a simple, economical and useful test to measure the lungs' capacity and functional status with devices called spirometers. In the PFT, the air volume that the lungs can take is measured by the air flow rate which is applied to the device during inspiration and then the airflow rate which is applied to the sensor during the expiration in breathing process. These airflow measurement rates are frequently used in differential diagnosis, following disease courses, and in the assessment treatment outcomes (Vaz Fragoso et al., 2010). Air volume entering and exiting the lungs is measured during challenging breathing, exercise and at rest; the obtained variability can be compared with the values obtained from the persons of the same age, sex, and height for diagnosis of the respiratory capacity.

When the lungs are full, FEV1 (Forced expiratory volume) is the maximum air volume during the breathing out in the first 1 second. The amount of air volume excreted from the lungs during forced expiratory maneuvers is called FVC (Forced Vital Capacity). FEV1, FVC, and FEV1/FVC (Tiffeneau-Pinelli index) values for respiratory disturbances have great importance in diagnosis, managing, determining the level and assessment processes of the respiratory disease as clinical data (Salvi & Barnes, 2009; Vaz Fragoso et al., 2010).

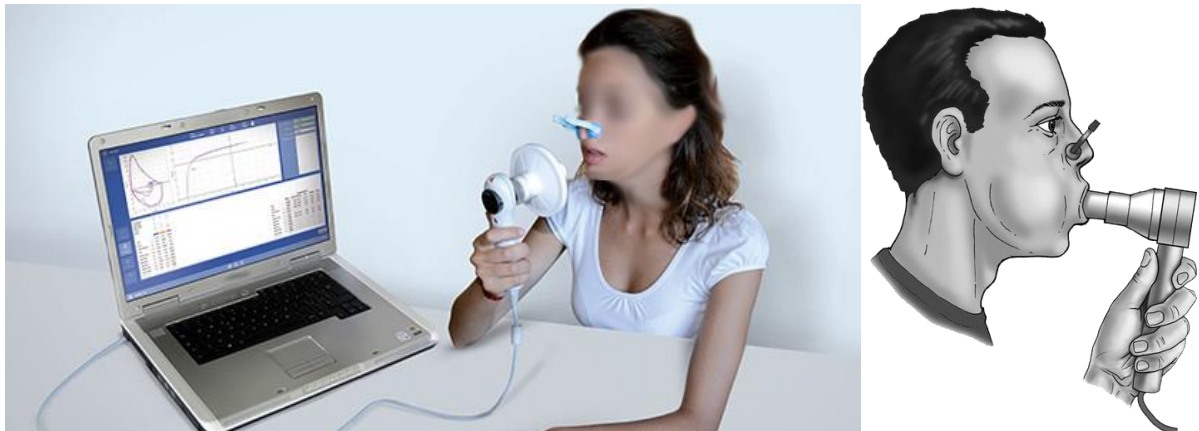

Figure 2. Pulmonary Function Test Scenario

The subjects were informed about inspiration and expiration in breathing and blowing procedure when before PFT. No breathing enhancer bronchodilator medication was used. The subject was allowed to perform all breathing operations by mouth using a nasal plug during PFT as seen in Figure 2. The disposable mouthpieces for the PFT device are used for every patient for preventing infectious diseases and are destroyed after measurements. After the mouthpiece has been placed in the mouth, attach importance to keep the lips are closed in order to avoid incorrect measurements. The subjects are asked for force, deep and fast breathing for at least 5 seconds. The obtained PFT variables and the test outcome are attached and imported to the designed CMI using the programmed PFT module. The COPD levels and symptoms for each level according to the results of PFT are described in (Roisin, 2016) and are given in Table 1.

Table 1. The COPD levels and symptoms for each level

| Levels | Symptoms |
|---|---|
| COPD0 : Under Risk | *PFT normal<br>*Have chronic symptoms (a bad or persistent cough, sputum) |
| COPD1 : Mild Level | *FEV1/FVC < %70<br>*FEV1 ≥ %80<br>*Have/not have chronic symptoms |
| COPD2 : Moderate Level | *FEV1/FVC < %70 |
| | * %50 < FEV1 < %80<br>* Have/not have chronic symptoms |
| COPD3: Severe Level | * FEV1/FVC < %70<br>* %30 < FEV1 < %50<br>* Have/not have chronic symptoms |
| COPD4: Very Severe Level | * FEV1/FVC < %70<br>* FEV1 < %30 or FEV1 < %50<br>* Chronic respiratory failure |

*Auscultation Scenario*

Auscultation is the general name given to the process of listening internal organs with the aid of a stethoscope. Auscultation is a commonly used diagnostic method for chest, heart, stomach and intestinal diseases. The most important characteristics of this diagnostic method are cheap, effective, easy to use and capable of detailed examination. Despite the continuous and rapid development of technology in chest diseases, the most commonly used and indispensable diagnostic method is still auscultation. Although there are standardized (CORSA) areas for chest auscultation (Celli et al., 2004; Sovijärvi et al., 2000), auscultation can be included if specific regions of the lung are deemed necessary. The thoracic auscultation areas used during physical examination include posterior-upper lung (L1-R1), posterior-middle lung (L2-R2), posterior-lower lung (L3-R3), posterior-costophrenic angle lung (L5-R5), and anterior-lower lung (L6-R6) for both Left (L) and Right (R) sides of the patients. The chest auscultation areas are demonstrated in Figure 3a for anterior side (chest wall) and in Figure 3b for posterior side (back).

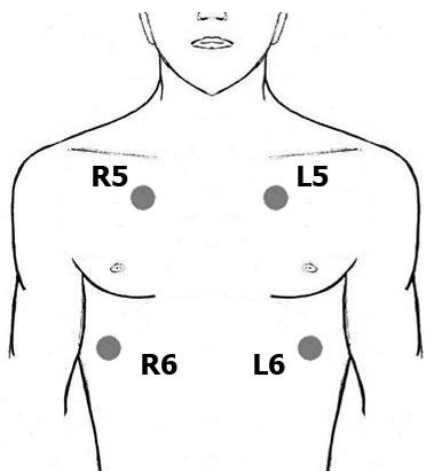
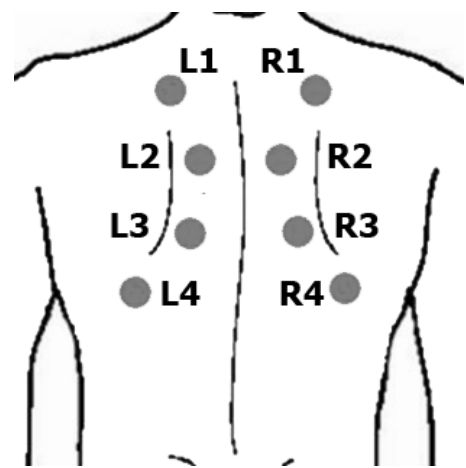

Figure 3a. Lung auscultation areas on anterior side

Figure 3b. Lung auscultation areas on posterior side

A consulting room isolated from external sounds, heated to adjust for the body temperature, well-lit, where the patient could feel comfortable was used during the auscultation procedure. The subjects were in seated position and auscultation sounds were recorded. The same auscultation points on left and right chest areas are auscultated synchronously using two digital stethoscopes by a pulmonologist clinician. The clinicians asked subjects not to speak during the recording for preventing non-natural artificial sounds from them. Once the recording is started, the patient is asked to cough once in the first 5 seconds and then to deeply breathe from mouth until the doctor has finished auscultation. Cardiac auscultation is performed in four different focuses: aortic (R7), pulmonic (L7), tricuspid (R8), and mitral (L8) areas. The heart auscultation areas are demonstrated in Figure 4.

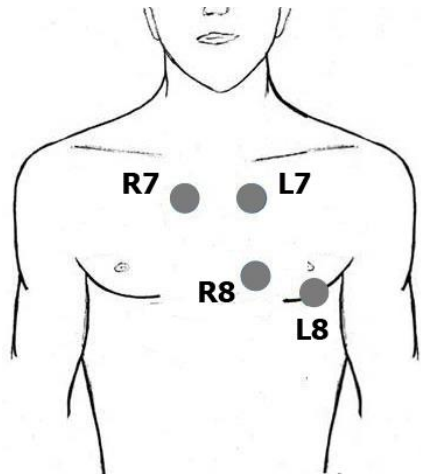

Figure 4. Heart auscultation areas

Breathing during heart auscultation causes to mix of breath sounds with heart sounds and heart sounds get away because of growing up in the rib cage. Hence, the pulmonologist clinicians ask from the subject coughing in the first 5 seconds when the recording started for the synchronization of the left and right side focused sounds and bating his breath to prevent mix-ups and losses on the auscultation sounds. The maximum peak which is occurred as a result of coughing in first 5 seconds was used to synchronize the auscultated sounds recorded in parallel using two digital stethoscopes. Recording was repeated when auscultation recordings did not meet the mentioned requirements of the scenario. Listening, point-based control, synchronization of two parallel auscultations and visualization of the lung and heart sounds which are recorded from the subjects can be performed using the designed CMI.

*Digitization of the Auscultation Sounds*

In literature, sensor based medical devices such as Piezoelectric microphone (Himeshima et al., 2012; Homs-Corbera et al., 2004; Matsutake et al., 2015; Umeki et al., 2015), electret microphone (Güler et al., 2005), LS-60 microphone (Waitman et al., 2000) electronic stethoscope (Dokur, 2009; Himeshima et al., 2012; Matsutake et al., 2015; Nakamura et al., 2016; Umeki et al., 2015) have been used while recording the auscultation sounds. It is imperative to use very sensitive devices to hear very short and hoarse lung and heart sounds such as crackles, wheezes and murmurs. Two of Littmann 3200 Electronic Stethoscopes were utilized in this study (Figure 5). This device provides an average reduction of 75% (-12dB) without removing the critical noisy body sounds. The latest technology filtering technology supplies three frequency modes (Bell, Diaphragm, and Extended) for auscultation of heart, lung and other internal body sounds. The device can transfer data to the PC in real time via Bluetooth technology or it can save 15 auspicious sounds to internal memory and can transfer bulk records into PC via Bluetooth ® communication. It acts as a diagnostic assistant, provides more effective use of telemedicine applications and increases the opportunity to teach auscultation sounds using the connection with the computer. It amplifies sounds up to 24x compared with acoustical stethoscopes. Response frequency intervals are Bell filter (20-200 Hz), Diaphragm filter (100-500 Hz), Extended filter (20-1000 Hz). The auscultation sound files are recorded with the computer zsa file extension. The file includes all auscultation parameters, subject information and sounds for each subject. The zsa files are converted to waveform audio file format (wav) with a sampling frequency of 4000 using the Littmann API.

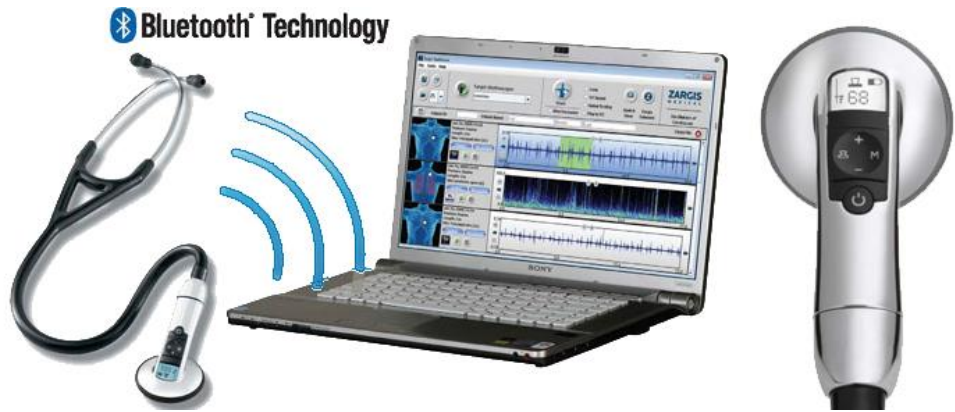

Figure 5. Littmann 3200 Digital Stethoscope

The database of the created CMI has been diversified with lung and heart auscultation sounds from patients with different stages of the COPD, lower respiratory tract disorders such as asthma and chronic bronchitis and healthy respiratory from subjects. Voluntary admittance was evaluated on a voluntary basis form with minimal information. The patients aged 38 to 68 are selected from different occupational groups, socio-economic status and genders for an accomplished analysis of the disorders. The subject population comprises of 13 of female and 64 of male. Distribution of diseases by gender is shown in Table 2.

Table 2. Diseases according to gender

| Diseases | # Records | Gender | |
|---|---|---|---|
| | | M | F |
| Asthma | 6 | 4 | 2 |
| COPD0 | 5 | 4 | 1 |
| COPD1 | 5 | 4 | 1 |
| COPD2 | 7 | 7 | - |
| COPD3 | 7 | 6 | 1 |
| COPD4 | 17 | 13 | 4 |
| Healthy | 30 | 26 | 4 |
| **Total** | 77 | 64 | 13 |

The most important reason for the COPD is the excessive use of tobacco products for many years. In the light of this information, the healthy auscultation records have been chosen among subjects who have never used cigarettes or tobacco products to obtain accurate and reliable data population. It has been noted that healthy subjects from different occupational and socioeconomic status have no diagnosed chronic lung history until now. Considering hereditary transmission of the COPD and asthma, the individuals whose close relative has asthma or the COPD were not included in the voluntary population.

A total number of 16-channel database is acquired with lung and heart auscultation sounds recorded from 8 basic foci for right and left sides on each subject. Each recorded sound is converted into 3 different frequency ranges: Bell, Diaphragm, and Extended filters. As a result, 48 (8x2x3) auscultation sounds were obtained for each subject. Considering the entire database, auscultation sound module has 3696 (77x48) of auscultated lung and heart sounds labelled for 5 of different respiratory diseases.

*Sound Synchronizer*

Synchronously recording from 2-channels for the left and right body areas are recorded for the RespiratoryDatabase@TR. The 2-channel auscultation sounds recording in parallel causes even at the millisecond level causes deviations. The synchronizer needs a triangulation point to form the beginning of two and more records. Pulling one or more trig points such as coughing peaks of two sounds will handle synchronizing the recorded auscultation lung and heart sounds.

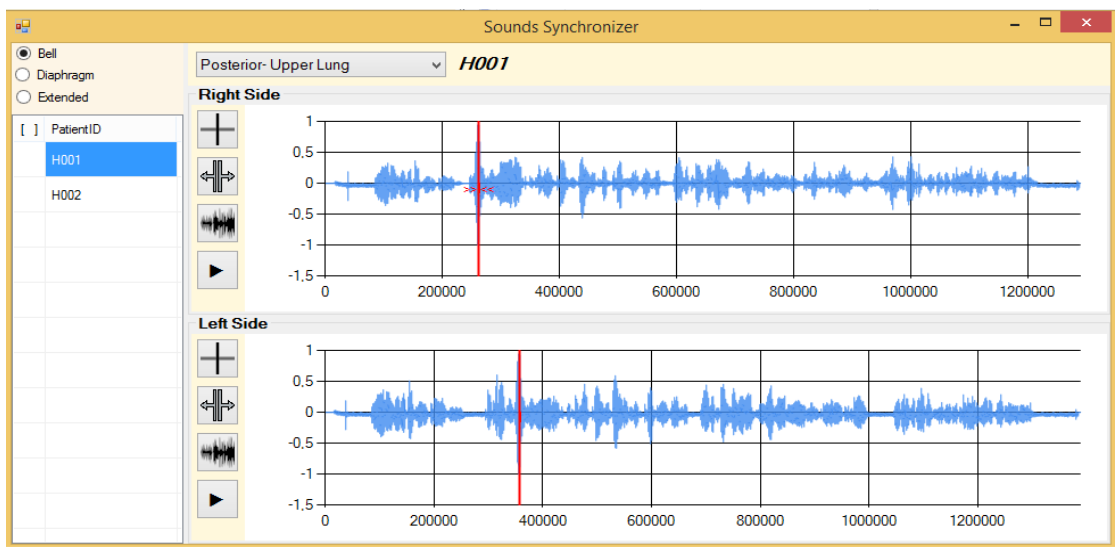

Figure 6. Sounds synchronization Module of the CMI

In this study, the patient was asked to cough in the first 5s before beginning breathing in. The coughing during the auscultation causes an unexpected excessive increase peaks in the sound signal. The peak points can be automatically detected in specific regions of the sounds by means of the designed CMI and marked with the annotation bar. In both left and right records, these peaks can be marked. When the synchronizer process is applied, The CMI module sets the marked positions as the beginning of the auscultated lung and heart sounds in interface module as seen in Figure 6.

*Sound Annotation*

The auscultation sounds can show different characteristics during breathing. Abnormal sounds such as crackle and wheezing sounds can be heard for short periods during both inspiration and expiration in certain auscultation sounds, they can be heard only in inspiration duration for some auscultation sounds. These time independent characteristics require regional labelling for the identic detection of murmur, crackle and wheezing sounds on the auscultation sounds.

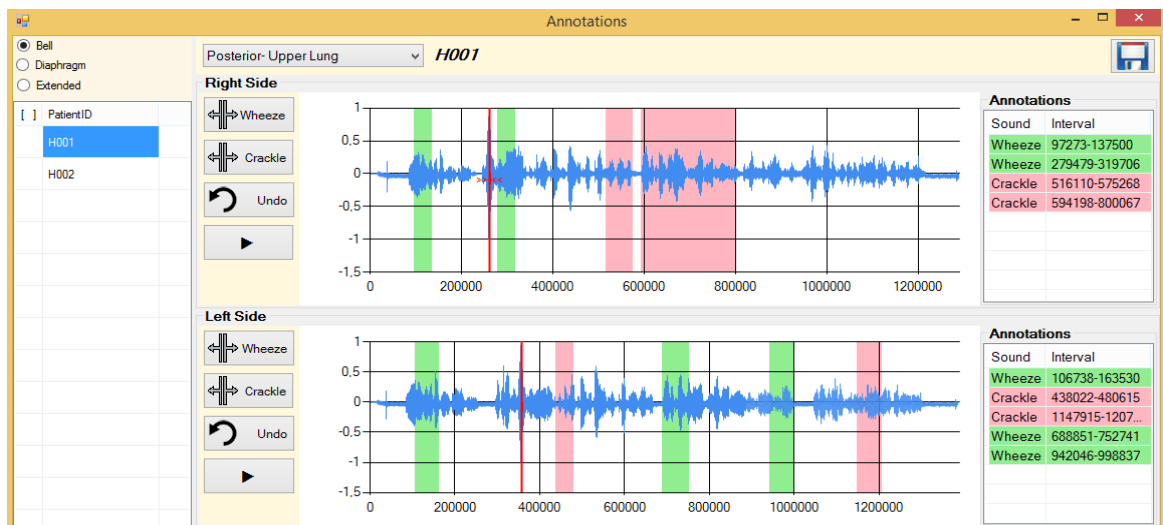

Figure 7. Auscultation Sounds Annotation Module of the CMI

The patient-based labelled auscultation sound region can be selected and can be analysed and listened on different frequency ranges by using the annotation module. On these recordings, annotation of adventitious wheezing and crackle sounds can be marked locally. Marked regions are recorded in the system according to audio-based annotation. The annotated regions can be listened directly, by eliminating normal respiratory sounds. The annotation module of the CMI is seen in Figure 7.

**Results**

Respiratory diseases are chronic and have seasonal reactions that must be constantly monitored and managed. The RespiratoryDatabase@TR is designed to track the condition of respiratory and pulmonary-cardiac diseases and to monitor the situation within the period. It features a unique auscultation and multimedia pulmonary database with the ability to analyse the lungs and the heart condition of auscultated sounds synchronized from 16 of different channels and a modular CMI that is easy to use for labelling and analysis on auscultation sounds and chest X-rays.

COPD has no return and treatment in the severe levels, and has an uncomfortable condition with manageable processes. Because of this, it is important early diagnosis and treatment to stop disease progression in early levels by keeping under control. The COPD and asthma are features that require constant control. The previous processes and attacks of the disease on the patient are very serious effects of treatment process. Because of the RespiratoryDatabase@TR is based on chronic disorders such as the COPD and asthma, it provides major advantages on allowing patients to easily compare the stages they have been through with the new processes of the patients being controlled, determining the extent the treatment processes successfully, and providing leading and formative help for different disciplines in the course of deciding the stage of treatment on respiratory diseases.

The RespiratoryDatabase@TR is a platform that comprises chest X-ray film, PFT measurements, SGRQ-C questionnaire answers, 16 channels of heart and lung auscultation sounds, and also a CMI to make ready respiratory multimedia files for computer based analysis and tracing the patients.

The PFT and chest X-rays are diagnostic tools that are performed according to certain standards worldwide. Lung auscultation is not a procedure linked to a specific standard in particular. But, literature has auscultation points that are used for the most clear and definitive diagnosis of the respiratory diseases. In this project, CORSA standardizations (Sovijärvi et al., 2000) were preferred for environmental and patient-based conditions such as breathing maneuvers. Aggregating according to certain standards on recording auscultation sounds and other multimedia data enables the RespiratoryDatabase@TR to be used and enhanced from all over the world. Having such a wide variety of multimedia content and a CMI with modular and effective algorithms provide ease of use has excellent potential and benefits in research and clinical detection.

Many studies in the literature have the characteristics of diagnosing pulmonary diseases or detecting abnormal sounds using only heart sounds or only lung sounds. The most important part of the medical health assessments which are difficult to accept; each patient is unique as naturally the symptoms are variable and the necessity of investigating as many symptoms as possible during the physical examination. The RespiratoryDatabase@TR allows chest X-ray, PFT values, SGQR-C status questioning, lung, and heart sounds to be used together, not just one of them for computerized-analysing on respiratory diseases. Chronic respiratory distress causes heart diseases in the following phases for the respiratory patients. The RespiratoryDatabase@TR ensures the assessment of heart and lung sounds together, and the identification of significant features for cardio-pulmonary diseases. Analysing possibility of the heart and lung sounds together provides opportunity on determining the association of chronic respiratory diseases with the heart diseases and whether diagnosing and assessments on the respiratory disorders such as the COPD and asthma using only heart auscultation sounds or other multimedia respiratory data. Parallel analysis of heart sounds and lung sounds from 16-channel and chest X-rays has provided researching opportunities to reach more meaningful features in the early diagnosis and diagnosis of cardio-pulmonary, cardiac and pulmonary diseases.

The chest X-rays, PFT variables, SGRQ-C questionnaire, and computerized auscultation sounds from so many channels allow for detailed examination of the patients and controlling the inner body over 16 auscultation areas. It has a higher potential for successful diagnosis because it has the ability to acquire higher quality and analysing capability than analogue auscultation. The auscultation module of the RespiratoryDatabase@TR has more permanent recording features for long time assessments and comparison and has the ability to be shared among doctors of different disciplines. The designed CMI has an easy access to all patient-based respiratory multimedia data; it can be repeated without any loss and may be duplicated. Digitized auscultation sounds can be analysed, quantified and visualized using signal processing methods easily during diagnosis and comparison the current phase with previous phases of the diseases. Computerized multimedia database features the modelling of new mathematical algorithms in the diagnosis of various pulmonary, cardiac and cardio-pulmonary diseases and stages of these diseases, approaches on image processing algorithms, researches on the development of classification algorithms, the establishment of embedded diagnostic tools and decision support systems.

Considering that medical image processing is one of the most studied fields in chest X-ray films, the RespiratoryDatabase@TR has qualities to be frequently referred to the determination of obstructions and abnormalities on respiratory diseases in chest X-rays, not only as auscultation sounds. The database and the CMI will be accessible online at www.respiratorydatabase.com.

**Acknowledgements**

This study is supported by Scientific and Technological Research Council of Turkish (TUBITAK-116E190). The authors express their thanks to TUBITAK for providing fully support.

**References**

Celli, B. R., MacNee, W., Agusti, A., Anzueto, A., Berg, B., Buist, A. S., Calverley, P.M.A., Chavannes, N., Dillard, T., Fahy, B., Fein, A., Heffner, J., Lareau, S., Meek, P., Martinez, F., McNicholas, W., Muris, J., Austegard, E., Pauwels, R., Rennard, S., Rossi, A., Siafakas, N., Tiep, B., Vestbo, J., Wouters, E., & ZuWallack, R. (2004). Standards for the diagnosis and treatment of patients with COPD: A summary of the ATS/ERS position paper. *European Respiratory Journal*. https://doi.org/10.1183/09031936.04.00014304

Decramer, M., Janssens, W., & Miravitlles, M. (2012). Chronic obstructive pulmonary disease. *Lancet*, *379*(9823), 1341–51. https://doi.org/10.1016/S0140-6736(11)60968-9

Dokur, Z. (2009). Respiratory sound classification by using an incremental supervised neural network. *Pattern Analysis and Applications*, *12*(4), 309–319. https://doi.org/10.1007/s10044-008-0125-y

Friis, B., Eiken, M., Hornsleth, A, & Jensen, A. (1990). Chest X-ray appearances in pneumonia and bronchiolitis. Correlation to virological diagnosis and secretory bacterial findings. *Acta Paediatrica Scandinavica*, *79*(2), 219–25. Retrieved from http://www.ncbi.nlm.nih.gov/pubmed/2321485

GBD 2015 Disease and Injury Incidence and Prevalence Collaborators, G. 2015 D. and I. I. and P. (2016). Global, regional, and national incidence, prevalence, and years lived with disability for 310 diseases and injuries, 1990-2015: a systematic analysis for the Global Burden of Disease Study 2015. *Lancet (London, England)*, *388*(10053), 1545–1602. https://doi.org/10.1016/S0140-6736(16)31678-6

Güler, E. Ç., Sankur, B., Kahya, Y. P., & Raudys, S. (2005). Two-stage classification of respiratory sound patterns. *Computers in Biology and Medicine*, *35*(1), 67–83. https://doi.org/10.1016/j.compbiomed.2003.11.001

Hederos, C.-A., Janson, S., Andersson, H., & Hedlin, G. (2004). Chest X-ray investigation in newly discovered asthma. *Pediatric Allergy and Immunology : Official Publication of the European Society of Pediatric Allergy and Immunology*, *15*(2), 163–5. https://doi.org/10.1046/j.1399-3038.2003.00098.x

Himeshima, M., Yamashita, M., Matsunaga, S., & Miyahara, S. (2012). Detection of abnormal lung sounds taking into account duration distribution for adventitious sounds. In *European Signal Processing Conference* (pp. 1821–1825).

Homs-Corbera, A., Fiz, J. A., Morera, J., & Jané, R. (2004). Time-Frequency Detection and Analysis of Wheezes during Forced Exhalation. *IEEE Transactions on Biomedical Engineering*, *51*(1), 182–186. https://doi.org/10.1109/TBME.2003.820359

Matsutake, S., Yamashita, M., & Matsunaga, S. (2015). Abnormal-respiration detection by considering correlation of observation of adventitious sounds. In *2015 23rd European Signal Processing Conference, EUSIPCO 2015* (pp. 634–638). https://doi.org/10.1109/EUSIPCO.2015.7362460

Nakamura, N., Yamashita, M., & Matsunaga, S. (2016). Detection of patients considering observation frequency of continuous and discontinuous adventitious sounds in lung sounds. *2016 38th Annual International Conference of the IEEE Engineering in Medicine and Biology Society (EMBC)*. https://doi.org/10.1109/EMBC.2016.7591472

Roisin RR. (2016). Chronic Obstructive Pulmonary Disease Updated 2010 Global Initiative for Chronic Obstructive Lung Disease. *Global Initiative for Chronic Obstructive Lung Disease. Inc*, 1–94. https://doi.org/10.1097/00008483-200207000-00004

Salvi, S. S., & Barnes, P. J. (2009). Chronic obstructive pulmonary disease in non-smokers. *The Lancet*. https://doi.org/10.1016/S0140-6736(09)61303-9

Sovijärvi, A. R. A., Vanderschoot, J., & Earis, J. E. (2000). Standardization of computerized respiratory sound analysis. *Eur Respir Rev*, *10*, 77–585.

Troosters, T., Casaburi, R., Gosselink, R., & Decramer, M. (2005). Pulmonary rehabilitation in chronic obstructive pulmonary disease. *American Journal of Respiratory and Critical Care Medicine*. https://doi.org/10.1164/rccm.200408-1109SO

Umeki, S., Yamashita, M., & Matsunaga, S. (2015). Classification between normal and abnormal lung sounds using unsupervised subject-adaptation. *2015 Asia-Pacific Signal and Information Processing Association Annual Summit and Conference (APSIPA)*. https://doi.org/10.1109/APSIPA.2015.7415506

Vaz Fragoso, C. A., Concato, J., McAvay, G., Van Ness, P. H., Rochester, C. L., Yaggi, H. K., & Gill, T. M. (2010). The ratio of FEV1 to FVC as a basis for establishing chronic obstructive pulmonary disease. *American Journal of Respiratory and Critical Care Medicine*, *181*(5), 446–451. https://doi.org/10.1164/rccm.200909-1366OC

Waitman, L. R., Clarkson, K. P., Barwise, J. A., & King, P. H. (2000). Representation and classification of breath sounds recorded in an intensive care setting using neural networks. *Journal of Clinical Monitoring and Computing*, *16*(2), 95–105. https://doi.org/10.1023/A:1009934112185

WHO. (n.d.). The top 10 causes of death. Retrieved August 7, 2016, from http://www.who.int/mediacentre/factsheets/fs310/en/